\documentclass[aps,preprintnumbers]{revtex4}
\usepackage{amssymb}
\usepackage{bm}
\usepackage{fancyhdr}
\usepackage[latin1]{inputenc}
\usepackage{amsmath}
\usepackage{empheq,amsmath}
\usepackage[english]{babel}
\usepackage{graphicx}

\def\M{\mathbb{M}}

\def\R{\mathbb{R}}

\def\M2{\R^{2 \times 2}}

\def\||{\parallel}

\def\sin{\mathrm{\; sin}\;}
\def\cos{\mathrm{\; cos}\;}

\def\qed{\vbox{\hbox to \textwidth{\hbox{} \rightline{\slshape q.e.d.} \hfil}}}

\begin{document}

\title{Fredholm Method for Podolsky Quantum Wave Function}
\author{Pedro Henrique Sales Girotto and Jorge Henrique Sales}
\affiliation{Universidade Estadual de Santa Cruz, DCET-PPGMC, 45 662-900, Ilhéus, BA,
Brazil}
\maketitle



\textit{\textbf{Abstract.}} \textit{In this paper we used the Fredholm
method in Schrödinger's integral equation in the investigation of the
scattering effect near the center of it between a stationary quantum wave
function and an electrostatic potential. Two potentials are studied one
Coulombian and the other Podolsky. The result shows the importance of the
proposal of Podolsky to regularize the effect near the scattering center in
the quantum wave function. Being that the coulombian potential presents with
strong variation in the amplitude of the wave after the scattering. In the
case of Podolsky's potential, this is corrected by adopting a constant that
removes this strong variation.}\newline
\textit{\textbf{Keywords:} Fredholm, Wave function, Podolsky}


\section{INTRODUTION}

The spontaneous breaking of chiral symmetry has a fundamental significance
in understanding the non-perturbative nature of hadron dynamics \cite{Nguyen}%
. It has been argued that the symmetry can be restored at sufficiently high
temperature. This breaking of symmetry around the center of the interaction (%
$x=0$), or scattering center is little investigated. In literature, it is
known as "End Point", where several papers seek to describe the energy
behavior in this scattering center \cite{Nguyen}.

The first vertex or central point to spread a quantum wave is given by
Coulombian interaction with an electrostatic potential that varies with the
inverse of the distance

\begin{equation}  \label{eq1}
V = \frac{1}{4\pi} \frac{Q^{2}}{r}
\end{equation}
where $Q$ is the central charge that scatter the wave and $r$ the distance
from the center any of the space. This potential shows that for the point in
the origin of the coordinate system $r=0$ implies in a singularity for the
potential energy. As a consequence, the solution to the stationary wave
function of the Schrödinger equation has a singularity at this point, in
this article this subject is seen in more detail.

The problem has originated from Maxwell's electromagnetic theory which has a
dependency $r^{-1}$ the Coulomb electrostatic potential for a point charge.
Thus, there is a divergence in both the energy as an electrostatic potential 
\cite{Sakurai}. A solution to this type of problem was proposed by Podolsky 
\cite{Podolsky42, Podolsky44, Podolsky48}, and consists of a generalization
of the theory of Electromagnetism in which is added a term of second order
in the derivatives of the electromagnetic field $A^{\mu }$, where $\mu
=0,1,2,3$ are the index of the space from Minkowski. In this way, the theory
that we have for this theory the Lagrangean of Podolsky.

\begin{equation}
\mathcal{L}_{0}=-\frac{1}{4}F_{\mu v}F^{\mu v}+\frac{a^{2}}{2}\partial
_{v}F^{\mu v}\partial ^{\alpha }F_{\mu \alpha }  \label{eq2}
\end{equation}%
where $F_{\mu v}=\partial _{v}A_{\mu }-\partial _{\mu }A_{v}$ and "a" is a
constant with length dimension. This Lagrangean generates a linear field
theory, which reduces to Maxwell's theory when $a=0$. This is evidently a
higher order theory since the equations of motion derived from Eq. (\ref{eq2}%
) contain quartic derivatives of the vector potential. Like Maxwell's
theory, Podolsky's theory also presents positive energy defined in the
electrostatic case, which, however, is finite for a point charge.

The latter result clearly shows that the force between two point charges is
no longer Coloumbian, a point which deserves closer examination. With this
theory of Podolsky the generalized electrostatic potential is obtained,
which then takes the form

\begin{equation}  \label{eq3}
V(r)= \frac{Q^2}{4\pi} . \frac{1-e^{-r/a}}{r}
\end{equation}
where $Q$ is the charge that generates the electrostatic field and $a$ is
the constant of Podolsky. The power is of the type Yukawa, with the
following properties: a finite value in the origin and converges to the
potential of Coulomb for $r >> a$, summarizing:

\begin{equation}  \label{eq4}
V(r) = 
\begin{cases}
\lim_{r \rightarrow 0} V(r) = \frac{Q^2}{4\pi a} \\ 
r >> a, V(r) = \frac{1}{4\pi} . \frac{Q^2}{r} \\ 
\lim_{r \rightarrow \infty} V(r) = 0%
\end{cases}%
\end{equation}

In this work, we show how a stationary quantum wave function, with Podolsky
potential, has no divergence for "a" different from zero. The method used is
the solution of the integral Schrödinger equation via Fredholm for the wave
function scattering by a Coloumbian (\ref{eq1}) and Podolsky (\ref{eq3})
potential.


\section{METHODOLOGY}

Integral equation occurs in a variety of applications, often being derived
from a differential equation. In possession of the integral equation the
Fredholm method is used with arbitrary kernels in the investigation of
possible singularities \cite{Fredholm}.

\subsection{Arbitrary kernels}

Be Fredholm equation of the second species:

\begin{equation}
u(x)=f(x)+\lambda \int_{a}^{b}K(x,t)u(t)dt  \label{eq5}
\end{equation}%
where $K(x,t)$ and $f(x)$ are known functions for $a\leq x\leq b$ and $a\leq
t\leq b$. This equation was solved by Fredholm in 1900 \cite{Fredholm},
using the artifice to replace the integral indicated in Eq. (\ref{eq5}) by
the corresponding sum of Riemann.

We divide the inter-value \textit{(a, b)} in \textit{n} inter-value equals 
\cite{Krasnov}:

\begin{equation}  \label{eq6}
\frac{b-a}{n} = \Delta x = \Delta t
\end{equation}
now, we will introduce the following notations: explanations

\begin{equation*}
\begin{matrix}
x_h(t_h) = a + h\Delta x(\Delta t) \\ 
f_i = f(x_i) \\ 
u_i = u(x_i(t_i)) \\ 
K_{pq} = K(x_p, t_q)%
\end{matrix}%
\end{equation*}
where $i,p,q = 1, 2, 3, \dots, n$

In this way, replacing the integral of Eq. (\ref{eq5}) by a summation, will
come:

\begin{equation}  \label{eq7}
u(x) = f(x) + \lambda \sum_{q=1}^{n} K(x,t_q)u_q\Delta t
\end{equation}
replacing in the above equation the variable $x$ by $x_p$, we obtain a
system of $n$ linear equations of the first degree of unknown functions $%
u_1, u_2, \dots ,u_n$. Therefore

\begin{equation}  \label{eq8}
\begin{matrix}
u_p = f_p + \lambda \sum_{q=1}^{n} K_{pq}u_q\Delta t &  &  &  &  & 
p=1,2,\dots,n%
\end{matrix}%
\end{equation}
to obtain the solution of the system given by Eq. (\ref{eq8}), lets do:

\begin{equation}
u_{p}=\sum_{q=1}^{n}\delta _{pq}u_{q}  \label{eq9}
\end{equation}%
where $\delta _{pq}$ is the Kronecker delta that, taking in Eq. (\ref{eq8}),
will give:

\begin{equation}
\begin{matrix}
\sum_{q=1}^{n}[\delta _{pq}-\lambda K_{pq}\Delta t]u_{q}=fp &  &  &  &  & 
p=1,2,\dots ,n%
\end{matrix}
\label{eq10}
\end{equation}%
using Cramer's rule \cite{Krasnov} to solve this system, we will have:

\begin{equation}
u_{q}=\frac{\Delta _{nq}(\lambda )}{\Delta _{n}(\lambda )}  \label{eq11}
\end{equation}%
where $\Delta _{n}(\lambda )$ is the determinant of the coefficients of the
dependent variables and $\Delta _{nq}(\lambda )$ is the determinant obtained
from this, in which the column is replaced of the coefficients of $u_{q}$ by
the column of independent terms $f_{p}$.

It was analyzed each of these determinants. The determinant $\Delta_n
(\lambda)$ It is given by:

\begin{equation}
\Delta _{n}(\lambda )=%
\begin{bmatrix}
1-\lambda K_{11}\Delta t & -\lambda K_{12}\Delta t & \dots  & -\lambda
K_{1n}\Delta t \\ 
-\lambda K_{21}\Delta t & 1-\lambda K_{22}\Delta t & \dots  & -\lambda
K_{2n}\Delta t \\ 
\vdots  &  & \vdots  &  \\ 
-\lambda K_{n1}\Delta t & -\lambda K_{n2}\Delta t & \dots  & 1-\lambda
K_{nn}\Delta t%
\end{bmatrix}
\label{eq12}
\end{equation}%
applying to the expression (\ref{eq12}) the decomposition formula of a
determinant \cite{Krasnov} will come:

\begin{equation}
\Delta _{n}(\lambda )=1-\frac{\lambda }{1!}\sum_{p_{1}=1}^{n}K_{p_{1}p_{1}}%
\Delta t+\frac{\lambda ^{2}}{2!}\sum_{p_{1}=1}^{n}\sum_{p_{2}=1}^{n}%
\begin{bmatrix}
K_{p_{1}p_{1}} & K_{p_{1}p_{2}} \\ 
K_{p_{2}p_{1}} & K_{p_{2}p_{2}}%
\end{bmatrix}%
(\Delta t)^{2}+\dots   \label{eq13}
\end{equation}%
\begin{equation*}
+(-1)^{n}\frac{\lambda ^{n}}{n!}\sum_{p_{1}=1}^{n}\sum_{p_{2}=1}^{n}\dots
\sum_{p_{n}=1}^{n}%
\begin{bmatrix}
K_{p_{1}p_{1}} & K_{p_{1}p_{2}} & \dots  & K_{p_{1}p_{n}} \\ 
K_{p_{2}p_{1}} & K_{p_{2}p_{2}} & \dots  & K_{p_{2}p_{n}} \\ 
&  & \vdots  &  \\ 
K_{p_{n}p_{1}} & K_{p_{n}p_{2}} & \dots  & K_{p_{n}p_{n}}%
\end{bmatrix}%
(\Delta t)^{n}
\end{equation*}%
before proceeding, we will introduce the following notation \cite{Krasnov}:

\begin{equation}
K\left( 
\begin{matrix}
x_{1} & x_{2} & \dots  & x_{n} \\ 
t_{1} & t_{2} & \dots  & t_{n}%
\end{matrix}%
\right) =%
\begin{bmatrix}
K(x_{1},t_{1}) & K(x_{1},t_{2}) & \dots  & K(x_{1},t_{n}) \\ 
K(x_{2},t_{1}) & K(x_{2},t_{2}) & \dots  & K(x_{2},t_{n}) \\ 
&  & \vdots  &  \\ 
K(x_{n},t_{1}) & K(x_{n},t_{2}) & \dots  & K(x_{n},t_{n})%
\end{bmatrix}
\label{eq14}
\end{equation}%
now we consider successively the terms of the second member of Eq. (\ref%
{eq13}). However, Riemann summations of this equation can be replaced by
integrals at the limit $n\rightarrow \infty $. Then, we will have,
respectively:

\begin{equation}
\sum_{p_{1}=1}^{n}K_{p_{1}p_{1}}\Delta t=\sum_{i=1}^{n}K(t_{1},t_{1})\Delta
t=\int_{a}^{b}K(t_{1},t_{1})dt  \label{eq15}
\end{equation}%
for the third term of Eq. (\ref{eq13})

\begin{equation}
\sum_{p_{1}=1}^{n}\sum_{p_{2}=1}^{n}%
\begin{bmatrix}
K_{p_{1}p_{1}} & K_{p_{1}p_{2}} \\ 
K_{p_{2}p_{1}} & K_{p_{2}p_{2}}%
\end{bmatrix}%
(\Delta t)^{2}=\int_{a}^{b}\int_{a}^{b}%
\begin{bmatrix}
K(t_{1},t_{1}) & K(t_{1},t_{2}) \\ 
K(t_{2},t_{1}) & K(t_{2},t_{2})%
\end{bmatrix}%
dt_{1}dt_{2}  \label{eq16}
\end{equation}%
and so successively. In this way, Equation (\ref{eq13}) will be:

\begin{equation}  \label{eq17}
\Delta_n (\lambda) = 1 + \sum_{n=1}^{\infty}(-1)^n\frac{\lambda^n}{n!}d_n
\end{equation}
where:

\begin{equation}
d_{n}=\int_{a}^{b}\int_{a}^{b}\dots \int_{a}^{b}K\left( 
\begin{matrix}
t_{1} & t_{2} & \dots  & t_{n} \\ 
t_{1} & t_{2} & \dots  & t_{n}%
\end{matrix}%
\right) dt_{1}dt_{2}\dots dt_{n}  \label{eq18}
\end{equation}%
with $K\left( 
\begin{matrix}
t_{i} \\ 
t_{i}%
\end{matrix}%
\right) $ given by Eq. (\ref{eq14}).

The value of the determinant $\Delta _{nq}(\lambda )$ of Eq. (\ref{eq11})
was obtained by Fredholm \cite{Fredholm}, with a calculation involving a lot
of algebraic manipulation, reason by why we will only present the result.
Like this:

\begin{equation}  \label{eq19}
\Delta_{nq}(\lambda) \equiv \Delta(x,t;\lambda) = K(x,t) +
\sum_{n=1}^{\infty}(-1)^n\frac{\lambda^n}{n!}d_n(x,t)
\end{equation}
where

\begin{equation}
d_{n}(x,t)=\int_{a}^{b}\int_{a}^{b}\dots \int_{a}^{b}K\left( 
\begin{matrix}
x & t_{1} & t_{2} & \dots  & t_{n} \\ 
t & t_{1} & t_{2} & \dots  & tn%
\end{matrix}%
\right) dt_{1}dt_{2}\dots dt_{n}  \label{eq20}
\end{equation}%
with $K\left( 
\begin{matrix}
x & t_{i} \\ 
t & t_{i}%
\end{matrix}%
\right) $ given by Eq. (\ref{eq14}).

Thus, according to Fredholm, the solution of Eq. (\ref{eq5}) will be given
by:

\begin{equation}
u(x)=f(x)+\lambda \int_{a}^{b}R(x,t;\lambda )f(t)dt  \label{eq21}
\end{equation}%
where the function $R(x,t;\lambda )$ is called the Fredholm resolvent Kernel
and defined by:

\begin{equation}
R(x,t;\lambda )=\frac{\Delta (x,t;\lambda )}{\Delta _{n}(\lambda )}
\label{eq22}
\end{equation}%
where $\Delta (x,t;\lambda )$ and $\Delta (\lambda )$ are given,
respectively, by Eq. (\ref{eq17}), (\ref{eq18}) and Eq. (\ref{eq19}), (\ref%
{eq20}).


\section{INTEGRAL EQUATION FOR AUTOFUNCTION}

We will only deal with the case of elastic scattering of particles whose
internal states do not change.

\subsection{Schrödinger integral equation}

The scattered particles move as free particles at a great distance from the
scattering center, the energy of their relative movement is always positive
and not quantized. Therefore, in the formulation of the problem of
scattering of a particle of mass m with positive relative energy E and in a
potential $V(\overrightarrow{r})$, it reduces to solve the Schrödinger
equation \cite{Sakurai}

\begin{equation}  \label{eq23}
H\psi(\overrightarrow{r}) = E \psi(\overrightarrow{r})
\end{equation}
Being:

\begin{equation}  \label{eq24}
H = -\frac{\hbar^2}{2m}\Delta\psi + V(\overrightarrow{r})
\end{equation}
Then of Eq. (\ref{eq23}) will stay:

\begin{equation}  \label{eq25}
(\Delta + k^2)\psi(\overrightarrow{r}) = \frac{2mV(\overrightarrow{r})}{%
\hbar^2}\psi(\overrightarrow{r})
\end{equation}
being

\begin{equation}  \label{eq26}
k^2 = \frac{2mE}{\hbar^2}
\end{equation}
the wave number of the incident particle with mass m and total energy E.

Out of the spreading region $(|\overrightarrow{r}| > d)$, it has been:

\begin{equation}  \label{eq27}
\begin{matrix}
V(\overrightarrow{r}) \neq 0 &  &  &  &  & |\overrightarrow{r}| \leq d%
\end{matrix}%
\end{equation}
therefore the Eq. (\ref{eq25}) will stay:

\begin{equation}
(\Delta +k^{2})\varphi (\overrightarrow{r})=0  \label{eq28}
\end{equation}%
whose solution equivalent \cite{Sakurai}:

\begin{equation}  \label{eq29}
\begin{matrix}
\varphi(\overrightarrow{r}) = \frac{1}{(2\pi)^{3/2}}e^{i\overrightarrow{k}.%
\overrightarrow{r}} &  &  &  &  & (\overrightarrow{p} = \hbar\overrightarrow{%
k})%
\end{matrix}%
\end{equation}

Now, let's solve Eq. (\ref{eq25}) for the scattering region. For this, we
will use the technique of the Green function \cite{Sales, Thibes}. Thus,
being the Green function for Eq. (\ref{eq25}) is given by:

\begin{equation}
(\Delta +k^{2})G(\overrightarrow{r},\overrightarrow{r}^{\prime })=\delta (%
\overrightarrow{r}-\overrightarrow{r}^{\prime })  \label{eq30}
\end{equation}%
the solution of that equation will be:

\begin{equation}
\psi (\overrightarrow{r})=\frac{2m}{\hbar ^{2}}\int G(\overrightarrow{r},%
\overrightarrow{r}^{\prime })V(\overrightarrow{r}^{\prime })\psi (%
\overrightarrow{r}^{\prime })d^{3}\overrightarrow{r}^{\prime }  \label{eq31}
\end{equation}%
In this way, the solution of the Schrödinger equation Eq. (\ref{eq25}) for
the whole space will be:

\begin{equation}
\psi (\overrightarrow{r})=\varphi (\overrightarrow{r})+\frac{2m}{\hbar ^{2}}%
\int G(\overrightarrow{r},\overrightarrow{r}^{\prime })V(\overrightarrow{r}%
^{\prime })\psi (\overrightarrow{r}^{\prime })d^{3}\overrightarrow{r}%
^{\prime }  \label{eq32}
\end{equation}%
Being the Green function is given by:

\begin{equation}
G(\overrightarrow{r},\overrightarrow{r}^{\prime })=\frac{1}{(2\pi )^{3}}\int 
\frac{e^{i\overrightarrow{K}.(\overrightarrow{r}-\overrightarrow{r}^{\prime
})}}{k^{2}-K^{2}}d^{3}K  \label{eq33}
\end{equation}%
For we make the integral indicated, initially we take the vector $%
\overrightarrow{r}-\overrightarrow{r}^{\prime }=\overrightarrow{\rho }$,
that is, in the direction of the polar axis of the space of $K$. Then:

\begin{equation}
\overrightarrow{K}.(\overrightarrow{r}-\overrightarrow{r}^{\prime })=K\rho 
\mathrm{\;\cos }(\theta )  \label{eq34}
\end{equation}%
On the other hand, being:

\begin{equation}
d^{3}K=K^{2}dK\mathrm{\;\sin }(\theta )d\theta d\phi   \label{eq35}
\end{equation}%
of Eq. (\ref{eq33}), will be:

\begin{equation}
G(\overrightarrow{r},\overrightarrow{r}^{\prime })=\frac{1}{(2\pi )^{3}}%
\int_{0}^{\infty }\int_{0}^{2\pi }\int_{0}^{\pi }\frac{e^{iK\rho \mathrm{%
\;\cos }(\theta )}}{k^{2}-K^{2}}K^{2}dk\mathrm{\;\sin }(\theta )d\theta
d\phi   \label{eq36}
\end{equation}%
Now, integrating into $\theta $ and $\phi $, we will have:

\begin{equation}
G(\overrightarrow{r},\overrightarrow{r}^{\prime })=\frac{1}{2\pi ^{2}\rho }%
\int_{0}^{\infty }\frac{\mathrm{\;\sin }(K\rho )KdK}{k^{2}-K^{2}}
\label{eq37}
\end{equation}%
How the integrand of Eq. (\ref{eq37}) is a pair function, then:

\begin{equation}
G(\overrightarrow{r},\overrightarrow{r}^{\prime })=\frac{1}{4\pi ^{2}\rho }%
\int_{-\infty }^{\infty }\frac{K\mathrm{\;}\sin \left( K\rho \right) dK}{%
k^{2}-K^{2}}  \label{eq38}
\end{equation}%
Using Euler's formula $e^{i\alpha }=\mathrm{\cos }(\alpha )+i\mathrm{\sin }%
(\alpha )$ and being made $K\rho =N$, of Eq. (\ref{eq38}) it becomes:

\begin{equation}  \label{eq39}
G(\overrightarrow{r}, \overrightarrow{r}^{\prime }) = -\frac{1}{8\pi^2\rho i}
\left[ \int_{-\infty}^{\infty}\frac{Ne^{iN}dN}{N^2 - \eta^2} -
\int_{-\infty}^{\infty}\frac{Ne^{-iN}dN}{N^2 - \eta^2} \right]
\end{equation}
where $\eta = k\rho$.

To make the integrals indicated in Eq. (\ref{eq39}), we will use the waste
method \cite{Zayats}. The contributions to these integrals come from the
zeros $q=\pm \eta $. The first integral is calculated using the
counterclockwise direction for a curve that circumvents the $+\eta $ pole.
Therefore, we will have:

\begin{equation}
\int_{-\infty }^{\infty }\frac{Ne^{iN}dN}{(N+\eta )(N-\eta )}=2\pi
iRes|_{N=\eta }=\pi ie^{i\eta }  \label{eq40}
\end{equation}%
Analogously, the second integral of Eq. (\ref{eq39}) is calculated using the
clockwise contour of the curve around the pole $-\eta $. Therefore, we will
have:

\begin{equation}  \label{eq41}
\int_{-\infty}^{\infty}\frac{Ne^{iN}dN}{(N+\eta)(N-\eta)} = (-2\pi i)
Res|_{N=-\eta} = -\pi i e^{i \eta}
\end{equation}
Taking the Eq. (\ref{eq40}) and (\ref{eq41}) into Eq. (\ref{eq39}), the
function of Green searched, it will be:

\begin{equation}  \label{eq42}
G(\overrightarrow{r}, \overrightarrow{r}^{\prime }) = -\frac{e^{ik|%
\overrightarrow{r} - \overrightarrow{r}^{\prime }|}}{4\pi|\overrightarrow{r}
- \overrightarrow{r}^{\prime }|}
\end{equation}
The choice of other a different contour than those used here would lead to a
term of the type $e^{-i\eta}$ in addition or in subtraction to the term $%
e^{+i\eta}$. Such term in $G(\overrightarrow{r}, \overrightarrow{r}^{\prime
})$ corresponds to an incident wave, which is contrary to Eq. (\ref{eq32}).

Therefore, the solution of Eq. (\ref{eq24}) will be obtained by tanking Eq. (%
\ref{eq42}) in Eq. (\ref{eq32}). So:

\begin{equation}
\psi (\overrightarrow{r})=\varphi (\overrightarrow{r})-\frac{2m\pi }{\hbar
^{2}}\int \frac{e^{ik|\overrightarrow{r}-\overrightarrow{r}^{\prime }|}}{|%
\overrightarrow{r}-\overrightarrow{r}^{\prime }|}V(\overrightarrow{r}%
^{\prime })\psi (\overrightarrow{r}^{\prime })d^{3}\overrightarrow{r}%
^{\prime }  \label{eq43}
\end{equation}


\section{RESULTS}

For an analysis of the behavior the stationary wave function scattering by a
coulombian potential, we go replacing $V(\overrightarrow{r}^{\prime })$ in (%
\ref{eq43}) by Coulomb's potential (\ref{eq1}) with load $Q=1$ in the
integral equation (\ref{eq43}), resulting in:

\begin{equation}
\psi (\overrightarrow{r})=\varphi (\overrightarrow{r})+\lambda
\int_{0}^{\infty }\frac{e^{ik|\overrightarrow{r}-\overrightarrow{r}^{\prime
}|}}{|\overrightarrow{r}-\overrightarrow{r}^{\prime }|}\frac{1}{%
\overrightarrow{r}^{\prime }}\psi (\overrightarrow{r}^{\prime })d^{3}%
\overrightarrow{r}^{\prime }.  \label{eq44}
\end{equation}%
Where $\lambda =-\frac{mQ^{2}}{2\hbar ^{2}}$, and to facilitate graphic
analysis is assumed $m=2$, $E=1/4$ and $\hbar =1$, this implies $\lambda =-1$
and $k=1$ in Eq. (\ref{eq26}).

The stationary wave function or self-function (\ref{eq44}) is an integral
equation of the Fredholm type (\ref{eq5}), whose solution is given by (\ref%
{eq21}). With the help of the mathematics program, was calculated the
determinants (\ref{eq17}) and (\ref{eq19}) which is soon replaced in (\ref%
{eq22}). For solution in a dimension $x$ we obtain:

\begin{equation}
\psi (x)=e^{ix}+\frac{e^{6i}\mathrm{\;\sin }(4)}{x}\frac{\lambda }{1+\lambda
\lbrack E_{i}(i)-E_{i}(5i)]}  \label{eq45}
\end{equation}%
where the function $E_{i}(t)$ is the integral exponential function given by

\begin{equation}
E_{i}(t)=\int_{t}^{\infty }\frac{e^{-u}}{u}du  \label{eq46}
\end{equation}

With all the considerations we have the graphical result, Fig. 1, for this
wave function with the Coulombian scattering potential:

\begin{figure}[h!]
\centering
\includegraphics{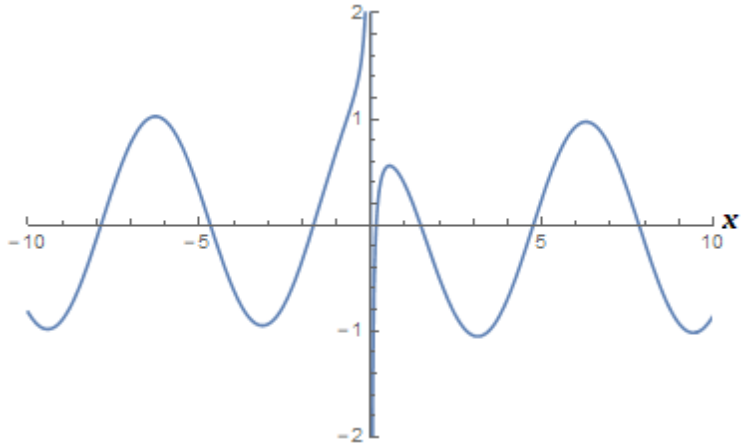}  
\caption{Wave function for Coulomb potential}
\label{fig1}
\end{figure}
Note that in the graph of Fig. \ref{fig1}, that the wave function (\ref{eq45}%
) as a function of distance $x$ has a behavior not defined in $x=0$. The
amplitude of this wave varies strongly between $x=0$ to $x=5$.

\begin{figure}[h!]
\centering
\includegraphics{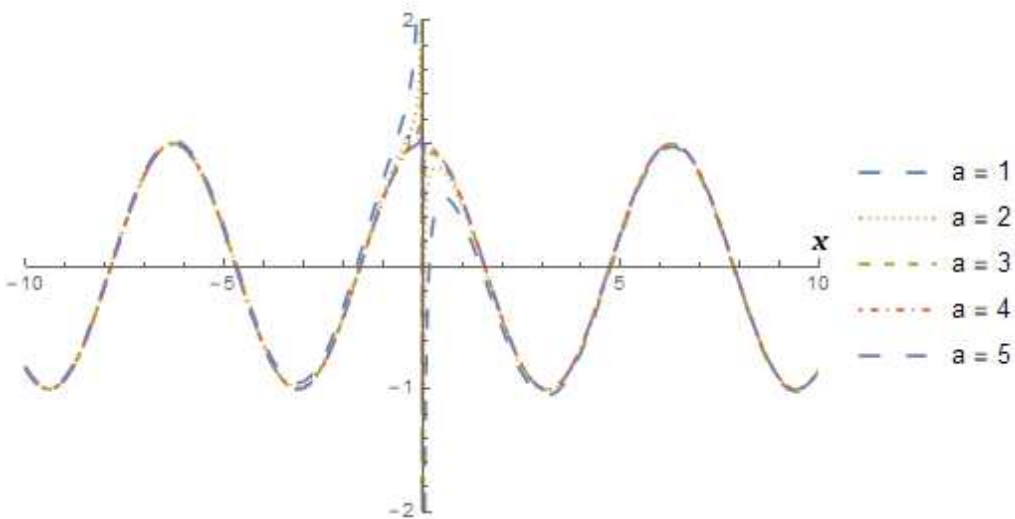}  
\caption{Wave function for Podolsky potential}
\label{fig2}
\end{figure}

The same procedure was used to calculate the steady-wave function for the
potential of Podolsky (\ref{eq3}), using values for constant $a=1,2,3,4,5$.
Resulting in the wave function

\begin{equation}  \label{eq47}
\psi(x) = e^{ix} + \frac{e^{2i-5a}(e^{4a}-e^{8i})}{x(a-2i)}\frac{\lambda}{1+%
\lambda[E_i(i-a)-E_i(5i-a)]}
\end{equation}
In Fig. \ref{fig2} is show the curves for each value of the Podolsky
constant "a". For $a = 1$ the wave function is close to the wave function
given by the coulombian potential, larger values of $a$ show a more stable
behavior indicating an amplitude tending to values between -1 to 1 before $x
= 5$.
\newpage

\begin{figure}[h!]
\centering
\includegraphics{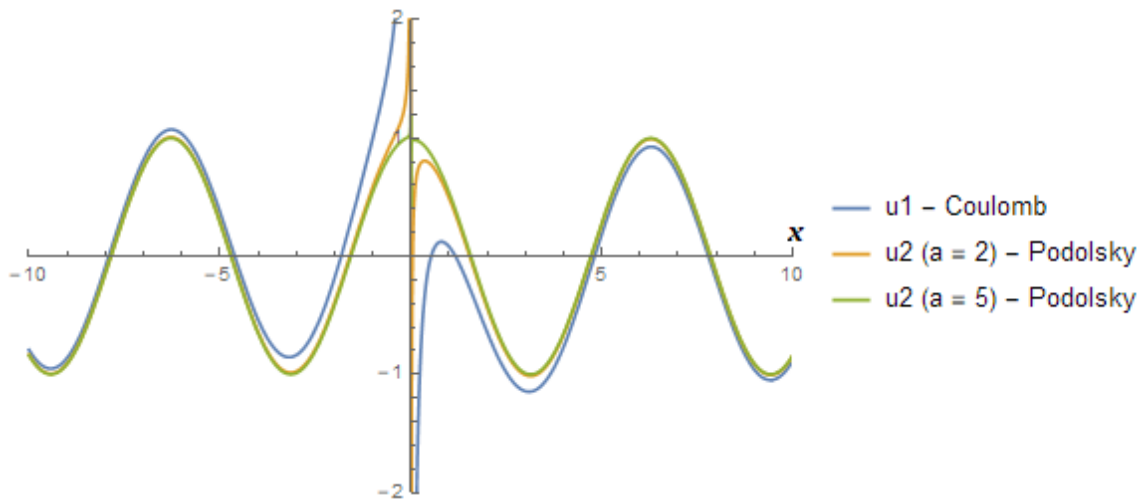}  
\caption{Wave functions scattered by Coulomb and Podolsky}
\label{fig3}
\end{figure}

Fig. \ref{fig3} shows the wave functions with Coulomb and Podolsky
scattering potential and it is observed that for $a=0$ it reproduces the
wave function with coulombian potential. Values for $a=2$ and $5$ indicate a
tendency to stabilize the wave with constant amplitude before the limit
value at the distance $x=5$, where $a=5$ is the best value. After this
limit, the waves of Podolsky practically propagate of similar form to that
of Coulomb.

\begin{figure}[h!]
\centering
\includegraphics[width=4in]{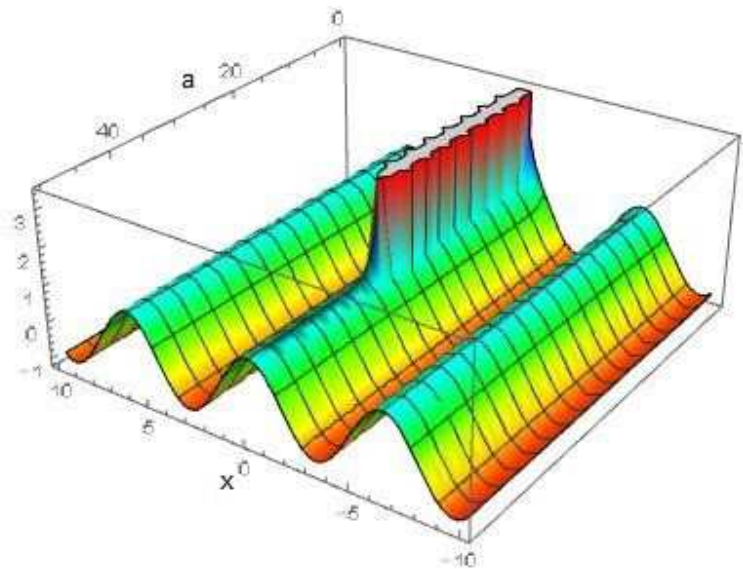}  
\caption{Scattered wave function by Podolsky potential}
\label{fig4}
\end{figure}

Fig. \ref{fig4} shows a graph for values of $x$ and $a$ varying. This
graphic evidences the importance of Constant $a$ the Podolsky. In fact,
constant $a$ the Podolsky, promotes a good behavior of the wave function
with values greater than zero. Since this constant $a$ is connected to the
distance dimension \cite{Podolsky48}, \cite{Thibes}, this explains why the
Podolsky wave function has a good behavior close to $x=0$ for values $a>0$.
\newpage


\section{CONCLUSION}

In this work was investigated the effect of the interaction at the center of
the scattering of a particle in the stationary wave function. The Schrödinger equation was used for two types of potential, one coulombian, and
other of Podolsky.

The technique used is the solutions of arbitrary kernels via integral
equation of Fredholm. The result of this work shows that in Fig. \ref{fig3}
the Podolsky constant $a = 0$ reproduces the wave function for the Coulomb
potential and an amplitude not defined at $x = 0$. For values, $a = 1$ and $5
$ the wave function for Podolsky's potential holds a constant amplitude at $
x = 0$. In Fig. \ref{fig4} can be observed from the graph that for $a > 40$
we do not have the singularity problem in the wave function at $x = 0$.
\newline

\textbf{Acknowledgment\newline
}JHS to CNPq and FAPESB for research grant and PHS to PPGMC-UESC.


\end{document}